\documentclass[aps,pre,epsf,twocolumn,showpacs]{revtex4}
\usepackage{amsmath}
\usepackage{epsfig}
\usepackage{amssymb}

\begin{document}

\title{Monte Carlo study of the interfacial adsorption of the Blume--Capel model}

\author{Nikolaos G. Fytas$^{1}$, Argyro Mainou$^{1}$, Panagiotis E. Theodorakis$^2$, and Anastasios Malakis$^{1,3}$}

\affiliation{$^1$ Applied Mathematics Research Centre, Coventry
University, Coventry CV1 5FB, United Kingdom}

\affiliation{$^2$ Institute of Physics, Polish Academy of Sciences,
Al. Lotnik\'ow 32/46, 02-668, Warsaw, Poland}

\affiliation{$^3$ Department of Physics, Section of Solid State Physics, University of Athens, Panepistimiopolis, GR 15784 Zografou, Greece}

\date{\today}

\begin{abstract}
We investigate the scaling of the interfacial adsorption of the
two-dimensional Blume--Capel model using Monte Carlo simulations.
In particular, we study the finite-size scaling behavior of the
interfacial adsorption of the pure model at both its first-- and
second--order transition regimes, as well as at the vicinity of the
tricritical point. Our analysis benefits from the currently
existing quite accurate estimates of the relevant
(tri)critical--point locations. In all studied cases, the numerical
results verify to a level of high accuracy the expected scenarios
derived from analytic free--energy scaling arguments. We also
investigate the size dependence of the interfacial adsorption
under the presence of quenched bond randomness at the originally
first--order transition regime (disorder--induced continuous
transition) and the relevant self--averaging properties of the system.
For this ex--first--order regime, where strong transient effects are shown to be
present, our findings support the scenario of a non--divergent scaling, similar
to that found in the original second--order transition regime of the pure
model.
\end{abstract}


\maketitle

Critical interfacial phenomena have been studied extensively over
the last decades, both experimentally and
theoretically~\cite{Abra,Diet,Bonn,Ral}. A well--known example is
wetting, where the macroscopically thick phase, \emph{e.g.}, the
fluid, is formed between the substrate and the other phase, say,
the gas. Liquid and gas are separated by the interface. An
interesting complication arises when one considers the possibility
of more than two phases. A third phase may be formed at the
interface between the two other phases. An experimental
realization is the two--component fluid system in equilibrium with
its vapor phase~\cite{Diet,Mold}. Both of the above scenarios may
be mimicked in statistical physics in a simplified fashion, by
either the two--state Ising model in wetting -- with the state
``+1'' representing, say, the fluid, and ``-1'' the gas -- or for
the case of a third phase via multi--state spin models, simply by
fixing distinct boundary states at the opposite sides of the
system. In this latter case, the formation of the third phase with
an excess of the non--boundary states has been called as
interfacial adsorption~\cite{Pesch,Huse,Fish}.

Throughout the years, various aspects of the interfacial
adsorption have been investigated via Monte Carlo methods and
density renormalization--group calculations on the basis of
specific multi--state spin models, namely Potts and Blume--Capel
models~\cite{Pesch,Huse,Yeo,Kroll,Yama91,Yama,Carlon,Alba,Fytas,Alba2,Alba3}.
Additional scaling and analytic arguments have been
presented~\cite{Huse,Kroll,Carlon,Lebo,Messa,Cardy}, though not
all of them have been confirmed numerically, due to the restricted
system sizes studied and, in some cases, the uncertainty in the
location of (tri)critical points. However, notable results in the
field include the determination of critical exponents and scaling
properties of the temperature and lattice--size dependencies, as
well as the clarification of the fundamental role of the type of
the bulk transition, with isotropic scaling holding at continuous
and tricritical bulk transitions, and anisotropic scaling at bulk
transitions of first--order type. More recently, a formulation of
the field theory of phase separation by Delfino and colleagues has
provided new insight into the
problem~\cite{Delfino12,Delfino13,Delfino14a,Delfino14b,Delfino15,Delfino16a,Delfino16b,Delfino18}
and, what is more, the role of randomness has been scrutinized on the basis of
the disordered Potts model~\cite{Monthus,Brener,Fytas2,Fytas3}.

Clearly the Potts model offers the unique advantage that if one
considers the system at its self--dual point, then, the
phase--transition temperatures between the ordered ferromagnetic
phase and the high--temperature disordered phase are known exactly
from self--duality for arbitrary values of the internal states $q$
and particular implementations of the randomness
distribution~\cite{Kinzel}. On the other hand, for the
Blume--Capel model, one relies upon the existing estimates for the
locations of (tri)critical and transition points and this may be a
source of systematic error when uncovering the scaling behavior of
the interfacial adsorption, as has already been underlined in the
literature~\cite{Kroll}. However, quite recently important
progress has been reported with respect to an accurate
reproduction of the phase diagram of the model for a wide range of
its critical
parameters~\cite{Silva,Malakis1,Malakis2,Kwak,Zierenberg,Fytas18},
thus motivating the current study. In the present
work we investigate the finite--size scaling behavior of the
interfacial adsorption of the two--dimensional square--lattice
Blume--Capel model, at both the continuous and first--order
transition regimes of its phase diagram, as well as at the
vicinity of the tricritical point. Furthermore, we study the
effect of quenched bond randomness on the interfacial adsorption
at the disorder--induced continuous transition. 
\begin{figure}[htbp]
\includegraphics*[width=8 cm]{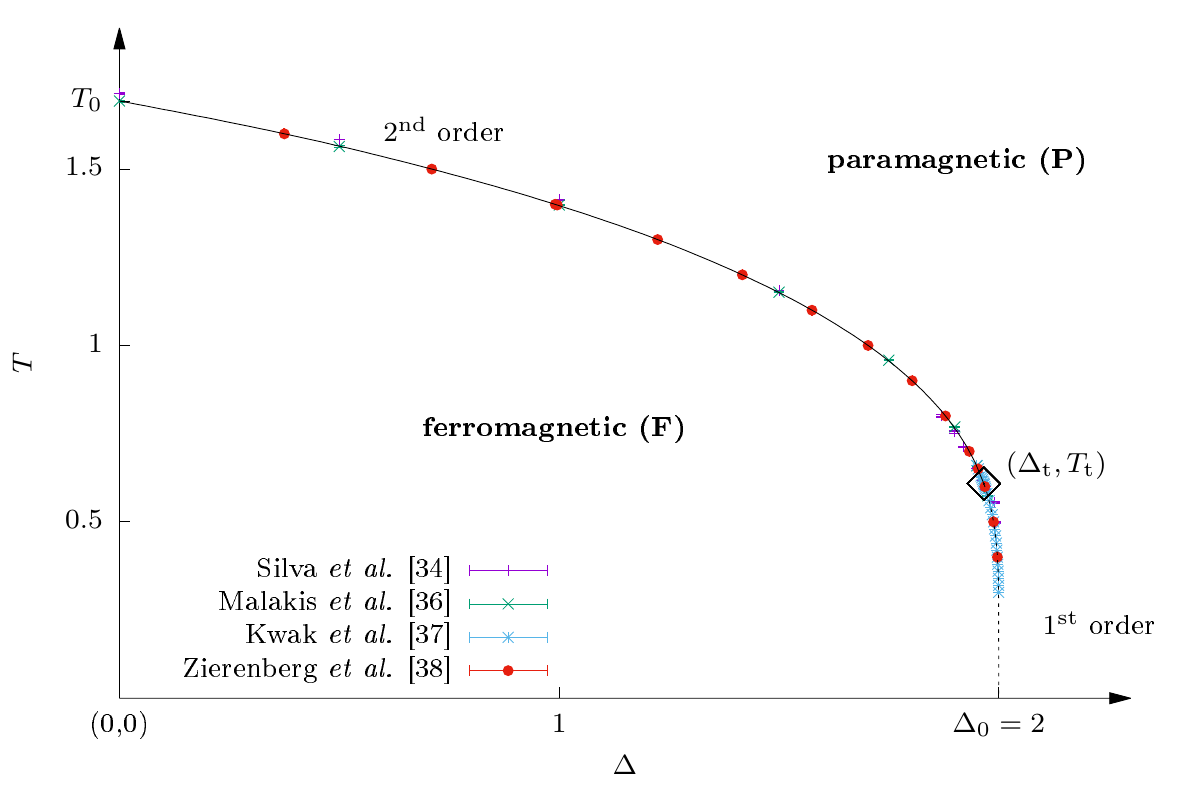} \caption{\label{fig:1}
Phase diagram of the square--lattice zero--field
Blume--Capel model in the $\Delta$ -- $T$ plane. The phase boundary
separates the ferromagnetic (F) phase from the paramagnetic (P)
phase. The solid line indicates continuous phase transitions and
the dotted line marks first--order phase transitions. The two lines
merge at the tricritical point $(\Delta_{\rm t}, T_{\rm t})$, as
highlighted by the black diamond. The data shown are selected
estimates from previous numerical studies. As usual, we have set $J = 1$ and
$k_{\rm B} = 1$ to fix the temperature scale.}
\end{figure}
Our discussion below follows the seminal works by Selke and
collaborators~\cite{Yeo,Kroll}, where the first Monte Carlo
results for the pure Blume--Capel model have been presented,
corroborated by analytical scaling arguments, which we will also
outline for the benefit of the reader in cases where a direct
comparison with the numerical data is possible. In a nutshell, the main objectives of the current work are as follows: For the pure case, previous numerical findings~\cite{Yeo,Kroll} based on less extensive simulations, are scrutinized, confirmed, and refined to a high--level of numerical accuracy, especially for the areas around the tricritical point and the first-order transition line in the $\Delta$ -- $T$ plane (as will be explicitly elaborated in the discussion of Figs.~\ref{fig:2} and \ref{fig:3} below). Completely new results are presented for the random case, an aspect that has not been previously considered in the relevant literature, where an intriguing crossover behavior, with a finite interfacial adsorption, at the
randomness--induced continuous transition is observed and explained. 

We consider the Blume--Capel model~\cite{Blume,Capel} defined by
the Hamiltonian
\begin{equation}\label{eq:1}
 \mathcal{H} = -J\sum_{\langle ij \rangle}S_{i}S_{j}+\Delta
 \sum_{i}S_{i}^2.
\end{equation}
The spin variable $S_{i}$ takes on the values $-1$, $0$, or $+1$,
$\langle ij \rangle$ indicates summation over nearest neighbors,
and $J>0$ is the ferromagnetic exchange interaction. The parameter
$\Delta$ denotes the crystal--field coupling and controls the
density of vacancies ($S_{i} = 0$). For $\Delta \rightarrow
-\infty$ vacancies are suppressed and the model becomes equivalent
to the Ising model. The phase diagram of the Blume--Capel model in
the crystal--field -- temperature plane consists of a boundary that
separates the ferromagnetic from the paramagnetic phase, see
Fig.~\ref{fig:1}. The ferromagnetic phase is characterized by an
ordered alignment of $\pm 1$ spins. The paramagnetic phase, on the
other hand, can be either a completely disordered arrangement at
high temperature or a $\pm1$--spin gas in a $0$--spin dominated
environment for low temperatures and high crystal fields. At high
temperatures and low crystal fields, the
ferromagnetic--paramagnetic transition is a continuous phase
transition in the Ising universality class, whereas at low
temperatures and high crystal fields the transition is of
first--order character~\cite{Blume,Capel}. The model is thus a
classical and paradigmatic example of a system with a tricritical
point $(\Delta_{\rm t},T_{\rm t})$~\cite{Lawrie}, where the two
segments of the phase boundary meet. At zero temperature, it is
clear that ferromagnetic order must prevail if its energy $zJ/2$
per spin (where $z$ is the coordination number, $z = 4$ in the
present case) exceeds that of the penalty $\Delta$ for having all
spins in the $\pm 1$ state. Hence the point $(\Delta_0=zJ/2,T=0)$
is on the phase boundary~\cite{Capel}. For zero crystal--field
$\Delta$, the transition temperature $T_0$ is not exactly known,
but well studied for a number of lattice geometries. A most recent
reproduction of the phase diagram of the model can be found in
Ref.~\cite{Zierenberg}, and is also given here in
Fig.~\ref{fig:1}, where a summary of results is presented from
various works in the literature.
A recent accurate estimation of the location of the tricritical point has been given in
Ref.~\cite{Kwak}: $(\Delta_{\rm t},T_{\rm t}) = (1.9660(1),
0.6080(1))$.

In order to study the interfacial adsorption, denoted hereafter as
$W$, and following the work of Selke and
collaborators~\cite{Yeo,Kroll} we shall employ special boundary
conditions, distinguishing the cases $[1:1]$ and $[1:-1]$ that
will favor the formation of an interface within the system. For
the case $[1:1]$, the spin variable is set at all boundary sites
equal to $1$, while for the case $[1:-1]$ the variable is set
equal to $1$ at one half of the boundary sites and to $-1$ at the
opposite half of the boundary sites. Typical equilibrium
configurations have verified that under these special boundary
conditions an excess of the non--boundary states, $S_{i} = 0$, is
generated at the interface (see for instance Fig. 1 in
Ref.~\cite{Yeo}). This phenomenon is described quantitatively by
the net adsorption per unit length of the interface, that is
defined with the help of the following mathematical
expression~\cite{Yeo}
\begin{equation}
\label{eq:2} W=\frac{1}{L} \sum_{i}
\left[\langle\delta_{0,S_{i}}\rangle_{[1:-1]}-\langle
\delta_{0,S_{i}}\rangle_{[1:1]}\right],
\end{equation}
where the angular brackets denote thermal averages and $L$ is the
linear dimension of the square lattice. The critical behavior of $W$ is
characterized by the critical exponents $x$ and $\omega$
via~\cite{Kroll}
\begin{equation}
\label{eq:3} W_{L} \sim L^{x}\;\;\; (T = T_{\rm c}),
\end{equation}
and
\begin{equation}
\label{eq:4} W_{t_{\rm c}} \sim t_{\rm c}^{-\omega} \;\;\; (L =
\infty),
\end{equation}
where $t_{\rm c} = (T_{\rm c} - T)/T_{\rm c}$ is the reduced
critical temperature for the standard case of a critical point.
Although the above Eqs.~(\ref{eq:3}) and (\ref{eq:4}) are
expressed for the usual case of continuous transitions, they can
be similarly generalized for the case of a tricritical point,
where $t_{\rm t} = (T_{\rm t} - T)/T_{\rm t}$, or for a
first--order phase transition, $t^{\ast} = (T^{\ast} -
T)/T^{\ast}$, where $T^{\ast}$ denotes now the corresponding
transition temperature.

\begin{figure}[htbp]
\includegraphics*[width=8 cm]{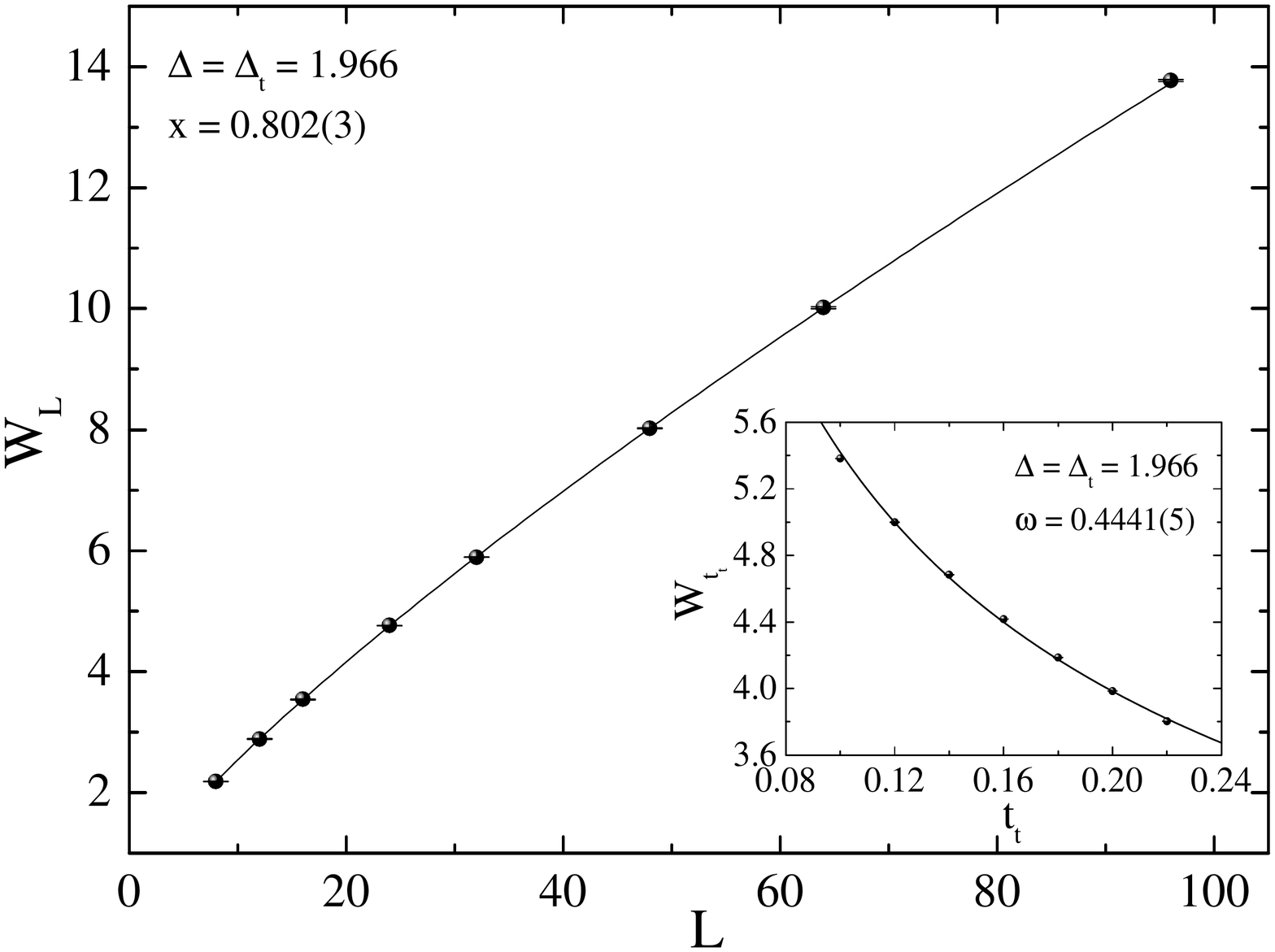} \caption{\label{fig:2}
Finite--size scaling of the interfacial adsorption $W_{L}$ (main
panel) and $W_{t_{\rm t}}$ (inset) at the tricritical point.}
\end{figure}

In the present work we have studied the interfacial properties of
the system at three values of the crystal--field coupling $\Delta$,
including both the first-- and second--order lines of the transition
but also the tricritical point of the phase diagram shown in
Fig.~\ref{fig:1}. We have considered the values
$\Delta = 1$ (second--order regime), $\Delta = \Delta_{\rm t} =
1.966$ (tricritical point), and $\Delta = 1.975$ (first--order
regime). The corresponding transition temperatures for the cases
$\Delta = 1$ and $\Delta = 1.975$ have been estimated to be
$T_{\rm c} = 1.398$ and $T^{\ast} = 0.574$,
respectively~\cite{Malakis2}, whereas for the case of the
tricritical point we have used the most recent estimate $T_{\rm t}
= 0.608$~\cite{Kwak}.
Additionally, for the case $\Delta = 1.975$ of the originally
first--order transition regime, we have also considered the
disordered version of the Hamiltonian~(\ref{eq:1}) by selecting
ferromagnetic couplings $J \rightarrow J_{ij}$ between
nearest--neighbor sites $i$ and $j$, to be either $J_1$, with
probability $p$, or $J_2$ with probability $1 - p$. In the case
$J_1 > J_2$, one has either strong or weak bonds. Then, the ratio
$r = J_2/J_1$ defines the disorder strength, where
$(J_1+J_2)/2=1$. Clearly, the value $r = 1$ corresponds to the
pure model. For the needs of the present work we fixed the ratio
$r = 0.6$, for which the critical temperature of the
disorder--induced continuous transition has been estimated to be
$T_{\rm c} = 0.626$~\cite{Malakis2}.

Our numerical protocol consists of canonical Monte Carlo
simulations, employing a combination of a Wolff single--cluster
update~\cite{Wolff1989} of the $\pm 1$ spins and a single--spin
flip Metropolis update that enables the necessary updates of the
vacancies $S_{i} = 0$~\cite{Blote95,hasen2010,malakis12}. We
adapted the relative frequencies of using the two updates to
optimize the performance and discarded the initial part of each time series to ensure
equilibration. Using this approach, we simulated for both versions
of the model and for all values of $\Delta$ system sizes in the
range $L = 8 - 96$, which, as will be shown below, is enough for a
safe estimation of the asymptotic behavior, in accordance with the
expected scaling arguments. 
\begin{figure}[htbp]
\includegraphics*[width=8 cm]{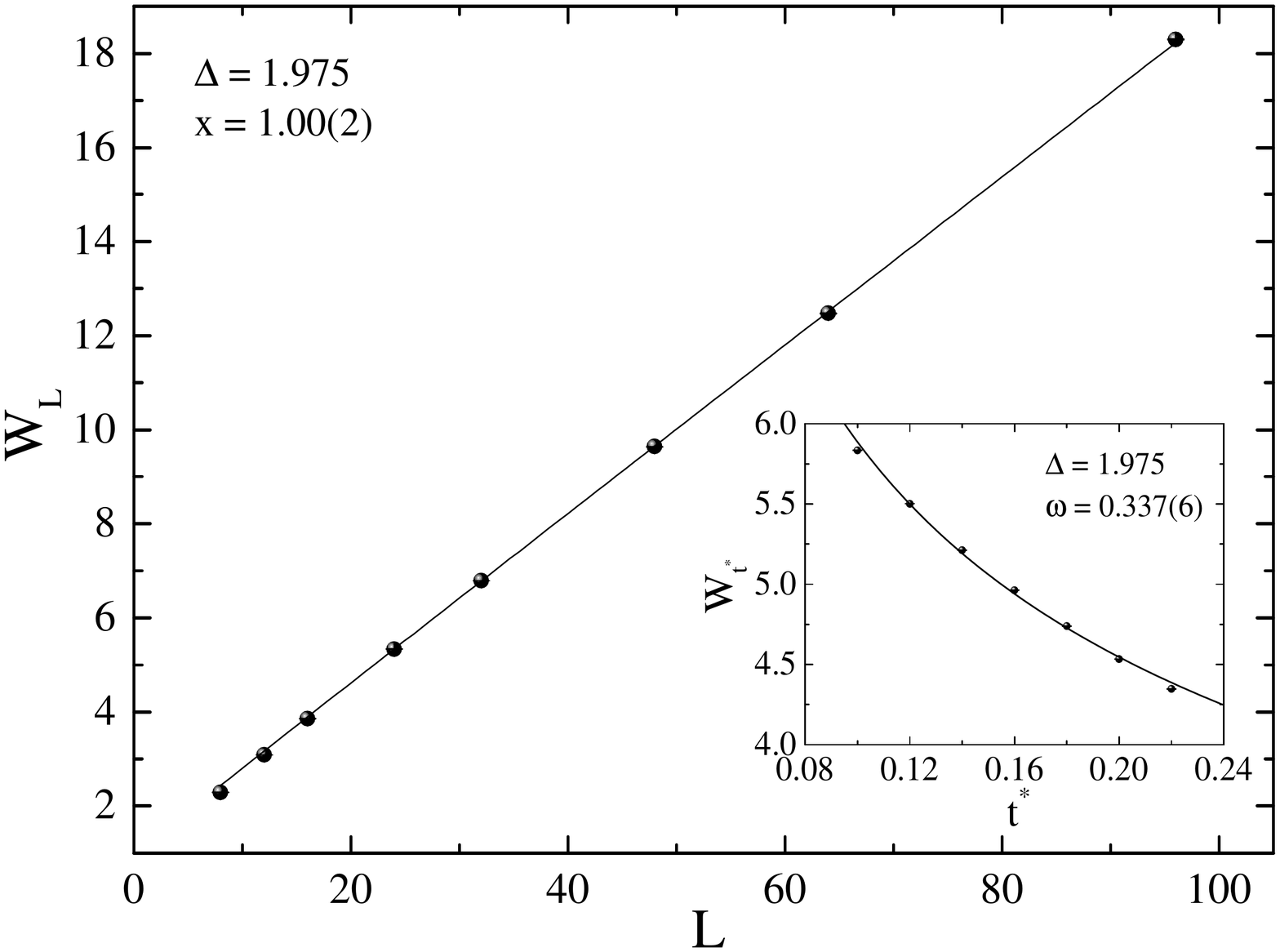} \caption{\label{fig:3}
Finite--size scaling of the interfacial adsorption $W_{L}$ (main
panel) and $W_{t^{\ast}}$ (inset) at the first--order transition
regime.}
\end{figure}
For the pure model we
performed several independent runs to increase statistical
accuracy, whereas for the disordered system an extensive averaging
over the disorder $[\ldots]$ has been undertaken, varying from
$5\times 10^3$ realizations for the smaller system sizes down to
$1\times 10^3$ for the larger sizes studied. For the disordered
case, error bars were computed from the sample--to--sample
fluctuations which in all cases were found to be larger than the
statistical errors of the single disorder realizations.

For the various cases of phase transitions in the Blume--Capel
model along the $\Delta$ -- $T$ plane, some very useful analytic
and scaling arguments for the interfacial adsorption have been
presented in the early work of Selke, Huse, and Kroll~\cite{Kroll}. 
In what follows, we shall only provide the main results of this discussion that are
also relevant for comparison with our numerical data; for more details we refer the reader to
Ref.~\cite{Kroll}. The main point in this description is the
reformulation of the interfacial adsorption $W$ with the help of
the interface tension $\sigma$. According to Ref.~\cite{Kroll},
using that $\langle\delta_{0,S_{i}}\rangle = 1 - \langle S_{i}^{2}
\rangle$, the interface adsorption may be written in the form $W =
(1/L) \sum_{i} \left[\langle S_{i}^{2} \rangle_{[1:1]}-\langle
S_{i}^{2} \rangle_{[1:-1]}\right]$. Denoting the total free energy
for $[1:1]$ boundary conditions by $F_{[1:1]}$ (similarly
$F_{[1:-1]}$ for $[1:-1]$), $W$ can then be expressed in terms of
the interface tension, $\sigma = (1/L)(F_{[1:1]} - F_{[1:-1]})$,
as $W = \beta^{-1}\partial \sigma / \partial \Delta$, where $\beta
= 1/(k_{\rm B}T)$.

The presentation of our finite--size scaling analysis starts with the most interesting cases referring to the vicinity of the tricritical point and the first--order transition regime. As already mentioned above, the location of the tricritical point of the Blume--Capel model is
known today with very good accuracy~\cite{Kwak}, thus removing one
source of error inherent in previous simulation
works~\cite{Yeo,Kroll}. According to the scaling arguments of
Ref.~\cite{Kroll} the exponents appearing in
Eqs.~(\ref{eq:3}) and (\ref{eq:4}) take on the values $x = 4
/ 5$ and $\omega = 4/9$, respectively, for the case of the
tricritical point. 
\begin{figure}[htbp]
\includegraphics*[width=8 cm]{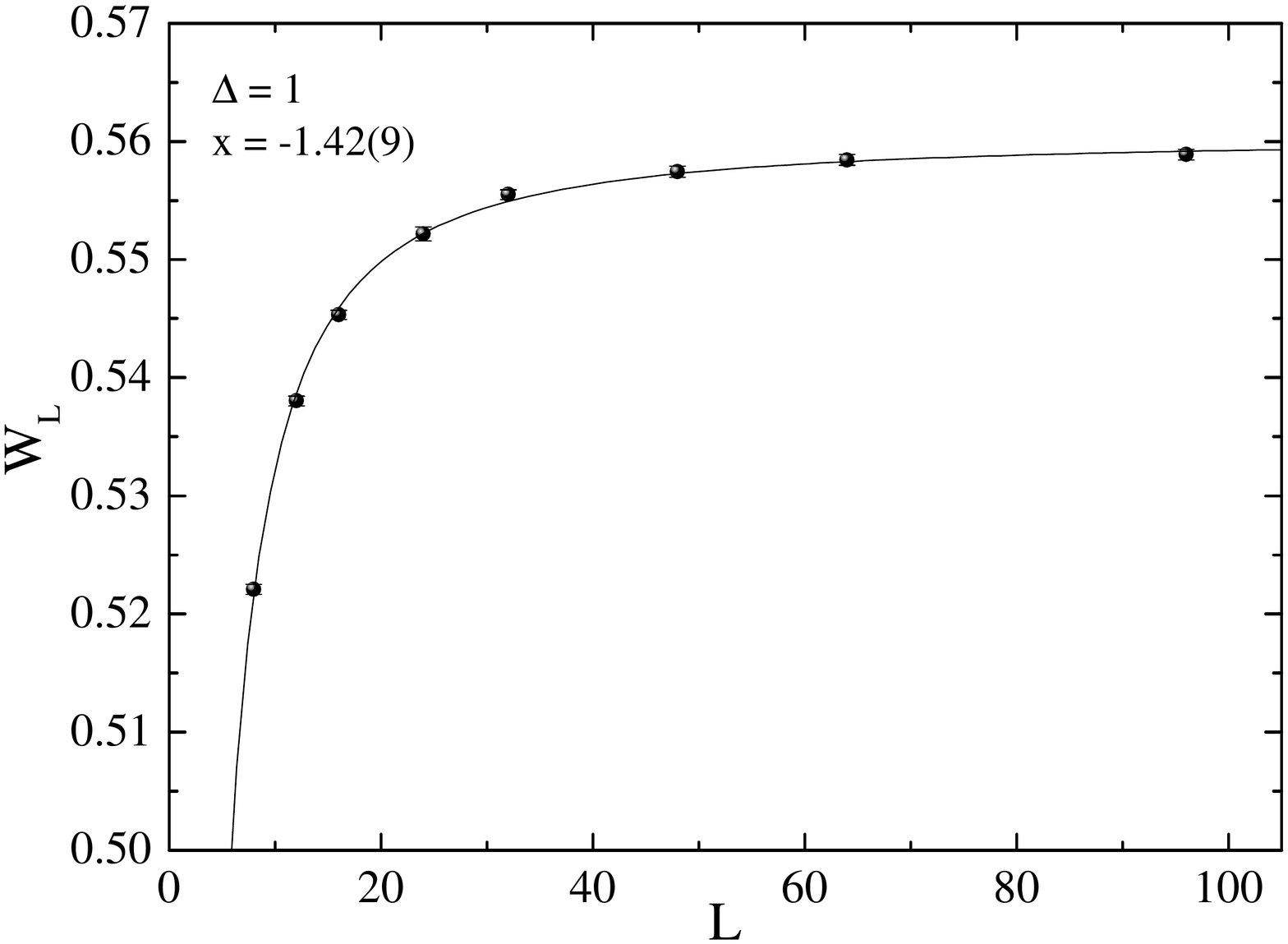} \caption{\label{fig:4}
Finite--size scaling of the interfacial adsorption $W_{L}$ at the
second--order transition regime.}
\end{figure}
In Fig.~\ref{fig:2} we present our numerical
data and the relevant scaling analysis for the interfacial
adsorption $W_{L}$ (main panel) and $W_{t_{\rm t}}$ (inset) at
$\Delta = \Delta_{\rm t} = 1.966$. Fits of the form~(\ref{eq:3})
and (\ref{eq:4}) shown by the solid lines in the main panel and
the corresponding inset respectively, provide us with the
estimates $x = 0.802(3)$ and $\omega = 0.4441(5)$, both fully
consistent with the expected values $x = 4 / 5$ and $\omega
= 4 / 9$.
We should point out here that the numerical estimation
of the exponent $x$ for the tricritical point has been reported as a quite
difficult task in the literature, due to the imprecise knowledge of the
tricritical coordinates (see Fig. 7 in Ref.~\cite{Yeo} where $\Delta_{\rm t} \approx 1.92(2)$) and the presence of strong finite--size effects for small system sizes (see Fig. 3 in Ref.~\cite{Kroll} where for the actual value of $\Delta_{\rm t} = 1.966$ an effective exponent of the order of $\sim 0.65$ is obtained). Both of these adversities have been satisfied in the present work,
leading to a clear verification of the scaling arguments presented in Ref.~\cite{Kroll}.

As it is well known, at the critical (and tricritical) points, the
singularities in the interfacial adsorption are induced by bulk
critical fluctuations.
On the other hand, at first--order phase
transitions there are no bulk critical fluctuations and the
divergence of $W$ arises from an interface delocalization
transition~\cite{Zia83}. In the latter case and for lattices of
square shapes a linear divergence of the form $W_{L} \sim L$ is
expected, \emph{i.e.}, $x = 1$~\cite{Kroll}.
Additionally, the
critical exponent $\omega$ appearing in Eq.~(\ref{eq:4}) is
expected to take the value $1/3$, as was originally found in the
case of interface unbinding~\cite{Smith82}, and further
generalized for first--order phases transitions in
two--dimensions~\cite{Yeo,Kroll,Zia83,Fisher84}. For the case of
the Blume--Capel model, the prediction $\omega = 1/3$ has been
numerically confirmed~\cite{Yeo,Kroll}, though the numerical data
for $W_{L}$ did not allow for an accurate estimation of the
exponent $x$. 
\begin{figure}[htbp]
\includegraphics*[width=8 cm]{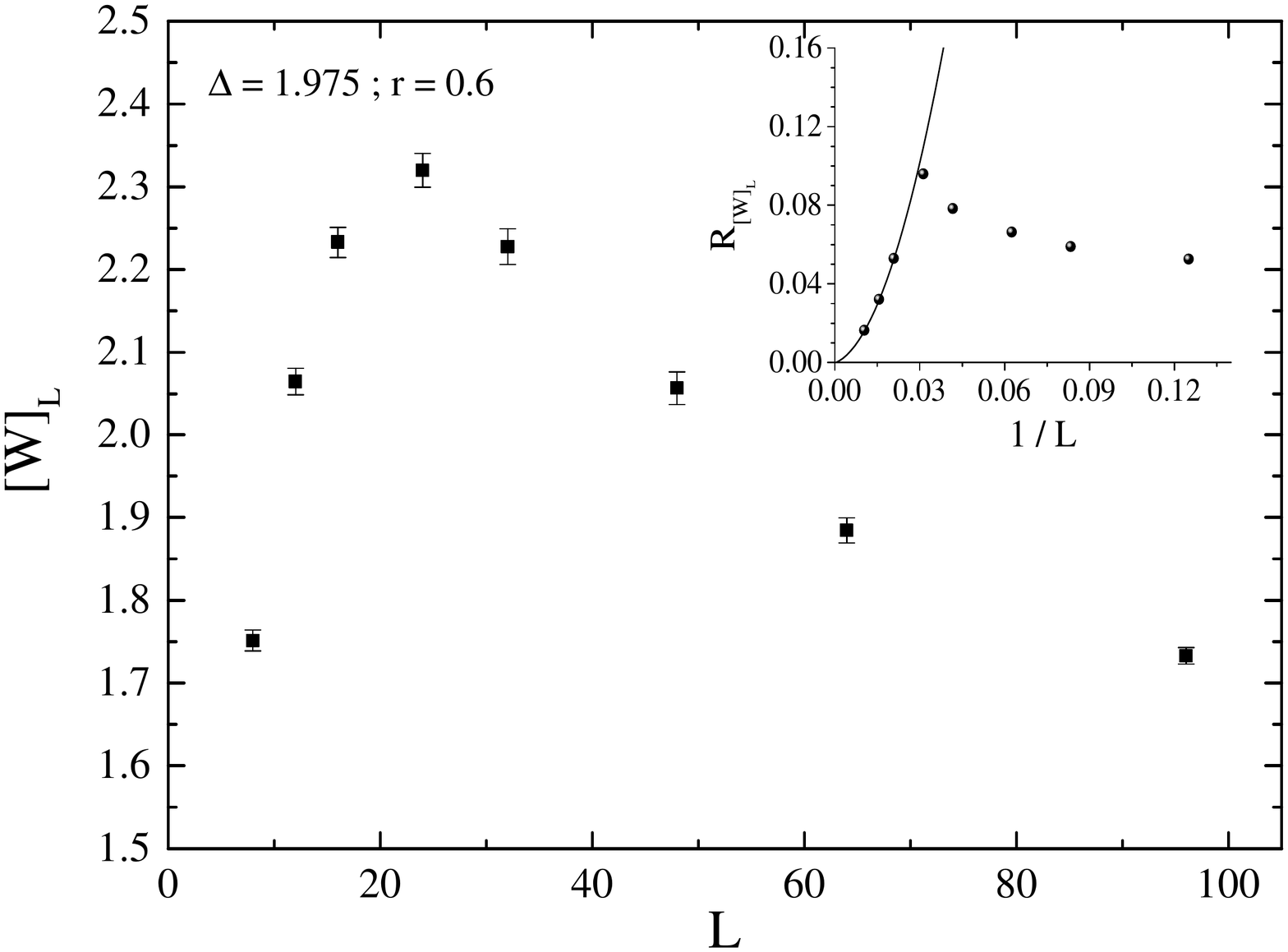} \caption{\label{fig:5}
Finite--size scaling of the disorder--averaged interfacial
adsorption $[W]_{L}$ of the random--bond Blume--Capel model at the
disorder--induced continuous transition. The inset illustrates the
relevant self--averaging properties in terms of the
relative--variance ratio $R_{[W]_{L}}$ as a function of the
inverse system size.}
\end{figure}
In particular, in Ref.~\cite{Yeo} a value $x = 0.7 \pm 0.05$ has been found that was subsequently explained as an apparent exponent due to strong metastability effects~\cite{Kroll}. To fill in the gap with the scaling analysis of $W_{L}$ at the first--order transition regime of the Blume--Capel model, we present in Fig.~\ref{fig:3} our numerical data for the interfacial adsorption
obtained at $\Delta = 1.975$. The fitting results using
the Eqs.~(\ref{eq:3}) and (\ref{eq:4}) as in Fig.~\ref{fig:3},
give $x = 1.00(2)$ and $\omega = 0.337(6)$, in excellent agreement
with the theoretical expectations $x = 1$ and $\omega = 1 / 3$.

For the spin--$1/2$ Ising model it is known that $\sigma \sim
t_{\rm c}$ for $t_{\rm c} \rightarrow 0^{+}$ at the critical
point. Given that the Blume--Capel model for $\Delta < \Delta_{\rm
t}$ belongs to the same universality class, we also expect a
similar statement to hold, where now $t_{\rm c}$ may be the
distance from the critical curve. Since $\Delta$ is a non--ordering
field~\cite{Riedel72}, as was also concluded in~\cite{Kroll}, $W
\sim \partial \sigma / \partial t_{\rm c} \sim$ const.
We present in Fig.~\ref{fig:4} the finite--size scaling behavior
of the interfacial adsorption $W_{L}$ for $\Delta = 1$. Indeed, a
simple power--law fit of the form $W_{L} = W_{\infty}+bL^{x}$
gives a negative exponent $x = -1.42(9)$ and a finite value of
$W_{\infty}$, thus a non--divergent behavior, in agreement with
the above arguments. Similar results have been presented in
Ref.~\cite{Yeo} for a few values of $\Delta$ in the second--order
transition regime but for smaller system sizes and are overall in
contrast to the Potts case, where a clear diverging behavior has
been observed in many relevant
works~\cite{Pesch,Huse,Fytas2,Fytas3}. This may be due to the
different geometric nature of the interfacial adsorption among the
two models, which in the present Blume--Capel model occurs in a
layer--like fashion as expected on the basis of single spin--flip
energy considerations, see Fig. 1 in Ref.~\cite{Yeo}, whereas in
Potts models a droplet--like adsorption of non--boundary states
takes place due to the energetic equivalence of all
states~\cite{Pesch}.

The last part of our work is dedicated to the study of the
interfacial adsorption under the presence of quenched bond
randomness at the originally first--order phase transition regime
of the phase diagram and particularly at the crystal--field value
$\Delta = 1.975$. Simulations have been performed for a single
value of the disorder strength, namely $r=0.6$, at the estimated
in Ref.~\cite{Malakis2} critical temperature $T_{\rm c} = 0.626$.
The numerical data for the disorder--averaged $[W]_{L}$ are shown
in the main panel of Fig.~\ref{fig:5}, where a very strong
saturation is observed~\cite{comment} and should be compared to
the diverging behavior of the corresponding pure system (see
Fig.~\ref{fig:2}). This result is in agreement with the
theoretical expectations discussed above for a non--divergent
behavior of $W$ in the case of continuous transitions for the
present model. Finally, in the inset of Fig.~\ref{fig:5} we
present the self--averaging properties of the system using the
relative--variance ratio $R_{[W]_{L}}=V_{[W]_{L}}/[W]_{L}^{2}$,
where $V_{[W]_{L}} = [W^2]_{L}-[W]_{L}^2$. The limiting value of
this ratio is characteristic of the self--averaging properties of
the system~\cite{WD95,AH96}. The solid line in the inset
illustrates a simple polynomial fit over the larger system sizes,
indicating the restoration of self--averaging at the thermodynamic
limit, given that $R_{[W]_{L}} \rightarrow 0$ as $L\rightarrow
\infty$. Similar results have been presented for the the case of
various random--bond Potts models in
two--dimensions~\cite{Fytas3}. Finally, it is worth noting that
the finite--size scaling behavior of both $[W]_{L}$ and
$R_{[W]_{L}}$ is affected by strong transient effects with a
crossover length--scale $L^{\ast} \approx 32$, where a turnaround
in the behavior sets off. This is consistent with previous
observations on the scaling behavior of the correlation length and
other thermodynamic observables of the system for the same range
of parameters~\cite{Fytas18}. Indeed, in Ref.~\cite{Fytas18} it has
been explicitly shown that $L \approx 32$ is the apparent size
where  the first--order characteristic signatures of the
transition disappear. Of course, we expect that the value of
$L^{\ast}$ depends on the disorder strength $r$ as well as on the
strength of the first--order transition and it would be
interesting to investigate the shift of this crossover
length--scale as a function of $\Delta$ and $r$. However this is a
task that goes beyond the scope of the present work.

To conclude, we have investigated the scaling aspects of the
interfacial adsorption of the two--dimensional Blume--Capel model
using a combined Monte Carlo scheme. We presented a
detailed analysis of the finite--size scaling behavior of the
interfacial adsorption of the pure model at both its first-- and
second--order transition regimes, as well as at the area of the
tricritical point, taking advantage of the current high--accuracy
estimates of (tri)critical--point locations. A dedicated part of our work regarding the scaling of the interfacial adsorption under the presence of quenched bond randomness at the originally
first--order transition regime (disorder--induced continuous
transition) revealed the scenario of a non--divergent scaling,
similar to that found in the original second--order transition
regime of the pure model. Overall, our results
and analysis nicely verified the predicted from analytic arguments
scaling scenarios of Ref.~\cite{Kroll}, overcoming the numerical difficulties 
highlighted in that seminal work.

\acknowledgments{N.~G.~F. would like to thank Prof. W. Selke for
many useful discussions on the topic of interfacial adsorption in 
multi--state spin models. This research has been supported by the National
Science Centre, Poland, under grant No.~2015/19/P/ST3/03541. This
project has received funding from the European Union's Horizon
2020 research and innovation programme under the Marie
Sk{\l}odowska--Curie grant agreement No. 665778. This research was
supported in part by PLGrid Infrastructure.}


\begin{thebibliography}{999}
\bibitem{comment} In fact we have not been able to fit the current data using a power-law fit of the form~(\ref{eq:3}), even after discarding the smaller system sizes.
\end{thebibliography}
\end{document}